\documentclass[12pt]{article}
\usepackage{epsfig}
\usepackage{amssymb}
\usepackage{amsmath}
\usepackage{graphicx}
\usepackage{appendix}

\newenvironment{mat}[1]%
{\left[ \begin{array}{#1} }%
{ \end{array} \right]}

\newlength{\espace}
\begin{document}

\title{\textbf{\textit{A Topological Approach to Scaling in Financial Data}}}

\author{\normalsize{Jean de Carufel \thanks{Corresponding author.
		Email: jdecarufel@apollosrc.com} , Martin Brooks , Michael Stieber, Paul Britton} \\
	\normalsize{Apollo System Research Corporation,} \\
	\normalsize{555 Legget Drive, Kanata, Ontario, Canada, K2K 2X3}} 

\date{\normalsize{October 2017}}

\maketitle

\begin{abstract}
	
There is a large body of work, built on tools developed in mathematics and physics, demonstrating that financial market prices exhibit self-similarity at different scales. In this paper, we explore the use of analytical topology to characterize financial price series.  While wavelet and Fourier transforms decompose a signal into sets of wavelets and power spectrum respectively,  the approach presented herein decomposes a time series into components  of its total variation.  This property is naturally suited for the analysis of scaling characteristics in fractals.    

\end{abstract}

\section{Introduction}

Financial market data provide some of the most complete historical measurements reflecting human social interaction.  Quantitative financial analysts submit these data to a wide range of analysis methods in an attempt to produce superior investment techniques, and in the process, often provide insight into the nature of interactions between humans as an ensemble.

As early as the 1930's, the scaling property of price series was studied using the analysis of repeatable patterns at various scales \cite{Elliot-1938}.  Early in his prolific career, Mandelbrot investigated the properties of price series \cite{Mandelbrot-1963} and following his work on establishing the theory of fractals, published  papers on the fractal nature and complexity of financial data \cite{Mandelbrot-1997} \cite{Mandelbrot-2001(1)} \cite{Mandelbrot-2001(2)} \cite{Mandelbrot-2001(3)}. Scaling relationships in price series have since been studied in various ways \cite{Muller-1990} \cite{Bouchaud-2000} \cite{DiMatteo-2003} \cite{Gencay-2001} \cite{Xu-2003}  and it has become widely accepted that self-similarity in prices is an emergent property of complex human interaction occurring in the market place.

The Fractal Market Hypothesis (FMH) \cite{Peters-1997} was established around the stylized fact that price series exhibit scaling characteristics.  The assumption that markets are instantaneously and completely \textit{efficient} (Efficient Market Hypothesis) was modified to introduce terms that support self-similarity at various time frames and price ranges.  According to the FMH, orderly markets are a result of the interaction between participants with a diverse set of investment horizons and views about current, and future asset valuations.  The scaling characteristics originate from the diversity of these market participants \cite{Kristoufek-2012}.  Consequently, it is plausible that scaling may be linked to the distribution of participants, and could be used as a measure of market condition. 

The determination of Hurst coefficients using the standard rescaled-range analysis \cite{Hurst-1951} was performed by Lo \cite{Lo-1991}. Di Matteo et al.\cite{DiMatteo-2003} then generalized the approach by using the q\textsuperscript{th}-moment of the return distribution to estimate a generalized Hurst coefficient.  This method  has the advantage of being less sensitive to tail events than Hurst's rescaled range method which relies on observed measurement extrema. A large number of scaling relationships were identified by Muller et al.\cite{Muller-1990}, and Dupuis and Olsen \cite{Dupuis-2012}, most of which were related to the count of price change events.  In recent years, the Fourier transform \cite{Machado-2102} and its fractional version \cite{Machado-2011} were also employed, as well as the Wavelet transform \cite{Bayraktar-2008}.  In all cases, the scaling characteristics of price series were demonstrated.

More recently, there has been an increased interest in topological data analysis (TDA).  Lemire et al. \cite{Lemire-2009}, Brooks \cite{Brooks-2016} and Seversky et al.\cite{seversky-2016} have used TDA for analyzing various data sets.  In particular, Brooks has developed a method consisting of decomposing the data into components of its \textit{total variation} and has applied it to the analysis of medical images \cite{Brooks-2016-2}. 

This paper presents results obtained from the application of TDA to price data.  The topological decomposition of the total variation is used to demonstrate the scaling property of financial instruments, both globally and as a time-varying property.   The paper is structured in the following way:  first we introduce the topological concepts required to understand the decomposition, we then apply the algorithm,  and finally we demonstrate the results of applying the proposed analysis method to price series for a foreign exchange market and a stock market index.  The estimated scaling characteristics are presented for each instrument, as well as its variation in time.   
 
\section{Topology Based Decomposition}  

\subsection{Total Variation for Single-Valued Real Functions}

Total variation refers to the total length of a curve over a closed subset of the domain.   For a single-valued real differentiable function $f(x)$ over an interval $x \in [a,b]$, the total variation is defined as 
\begin{equation}
TV(a,b) = \int_a^b 	\left| \frac{df(x)}{dx} \right| dx
\end{equation}
where $\frac{df(x)}{dx}$ is Riemann integrable.

For a prices series $p=\{p_0, p_1 , ..., p_k\}$ which consists of $k$ discrete samples, the total variation is obtained by summing up all price variations (i.e. \textit{coastline}), 
\begin{equation}
TV(p) = \sum_{i=1}^k \lvert r_i \lvert 
\end{equation}
where $r_i = \lvert p_i-p_{i-1} \lvert $ is the price return at time $i$.

\subsection{Persistent Pairs in Single-Valued Real Functions}

\textit{Persistence} is a concept in algebraic topology, well described by Edelsbrunner \cite{Edelsbrunner-2008} for single variable functions and simplical homologies. For this paper, we concentrate solely on the definition for single variable functions and introduce the persistent extrema pairing process for discrete price samples.  

\begin{figure}
	\centering 
	\includegraphics[width=14cm, height=9cm]{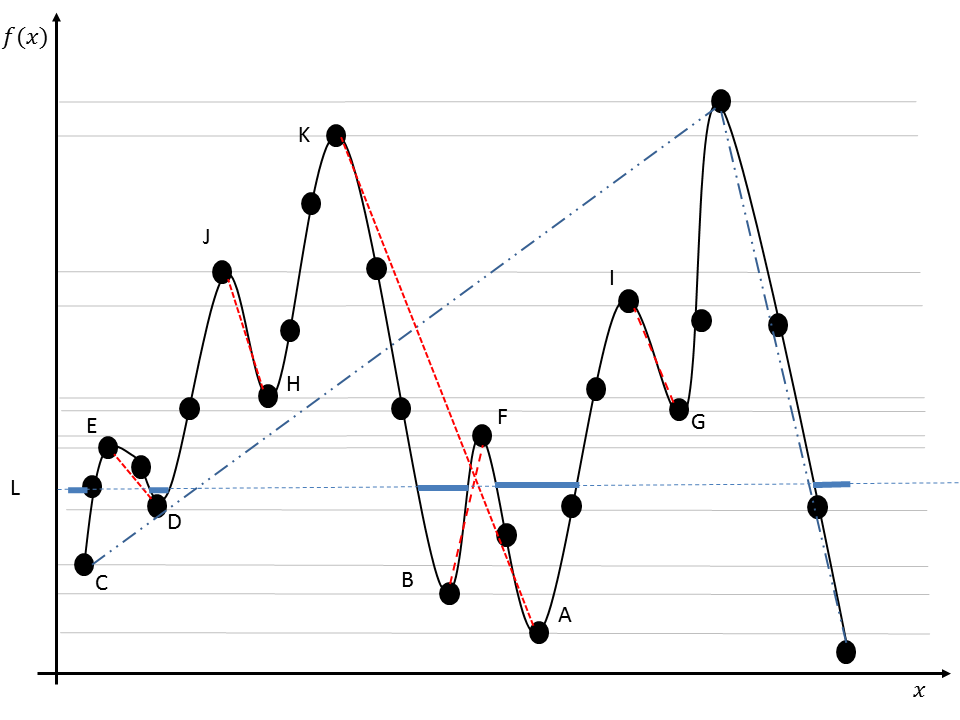}
	\caption{Persistent Pairing Process and Persistent Pairs}
	\label{PersistentPairs}
\end{figure}

As per Edelsbrunner \cite{Edelsbrunner-2008}, persistent pairs can be obtained by progressively increasing the detection level $L$ from the lowest value to the largest value in the set.  For a differential function $f:X \rightarrow Y$, the sublevel set $\Re(x) =\{\eta \in X | f(\eta)\le f(x)\}$ is the set of disjoint components representing continuous pieces of the domain X.  This is depicted in Figure \ref{PersistentPairs} where the bold horizontal line at level $f(x)=L$ represents $\Re(L)$. As $L$ increases, a new contiguous component is created every time a local minimum is reached.  Whenever a local maximum is reached, two components join into one.  At that point, a persistent pair is formed, consisting of the local maximum and the highest local minimum in the newly formed joined component. Referring to Figure \ref{PersistentPairs} and starting from the bottom, we first encounter local minimum A, then local minimum B, then local minimum C and finally local minimum D.  The process continues to the local maximum E, at which point two components merge into one.  The local maximum E creates a first persistent pair with local minimum D.  As the level progresses further, local maximum F is reached, which is paired with local minimum B to create a second persistent pair FB.  The process continues until no more pairing is possible. In the event that a local extrema has the same value for multiple samples, the persistent pair may not be uniquely defined.  In that case, either symbolic perturbation is used to \textit{perturb} the critical points and get an approximate solution, or the earliest extrema in the time-series is selected for persistence.  The latter approach is used here. The algorithm presented in Appendix \ref{algo} was taken from Lemire \cite{Lemire-2009} \cite{Lemire-2005}.  It extracts persistent movements on the fly, as new data points are provided.  The output is composed of a list of persistent pairs as well as a top structure.

In Figure \ref{PersistentPairs}, the dashed lines are the persistent movements while the dash-dot lines are the non-persistent movements defining the \textit{top structure} $T$ made up of extrema that are not yet persistent.  The price movements described by the persistent pair extrema are called \textit{persistent movements}.  They are price variations that do not contain a reversal larger than their own size, hence the term persistent movements.  The difference in value between extrema of a persistent pair is called the persistent movement size $\mu_j$.  It defines the price variation amplitude associated with the persistent movement. The total variation provided by a persistent movement $j$ is in fact twice its size $\mu_j$.  

A property of persistent movements is that the total variation is maintained, such that for a prices series $p=\{p_0, p_1 , ..., p_k\}$, 
\begin{equation}
TV(p) =\sum_{i=1}^k \lvert r_i \lvert = TV(T)+\sum_{j=1}^{n_p} 2 \mu_j
\end{equation}
where $r_i = \lvert p_i-p_{i-1} \lvert$, $TV(T)$ is the total variation associated with the top structure $T$ and $n_p$ is the number of persistent pairs.

\subsection{Persistent Movement Spectrum}

In a way similar to Fourier analysis where the spectrum indicates the signal energy at given frequencies, it is possible to define the persistent movement spectrum which indicates the contribution of each persistent movements of size $m_j$ to the total variation.  For price series, the set of discrete sizes $m_j$ is well defined since price data exists only at discrete values specified by the smallest price variation of the asset.  So the spectrum simply consists of counting the number of persistent movements $n_j$ of size $m_j$.  Like the Fourier transform where trends need to be removed and the data windowed to prevent distortion, the persistent movement spectrum applies only to the persistent price movements, and does not include the top structure.  The persistent movement spectrum is defined as 
\begin{equation}
S(m_j)=2 n_j m_j
\end{equation}
where $n_j$ is the number of persistent movements with size $m_j$.   

\subsection{Scaling Property of the Persistent Movement Spectrum}

We will demonstrate later that the persistent movement spectrum of a price series is in fact well represented by a power-law.  This indicates that the contribution to the total variation exhibits self-similarity with respect to the persistent movement size.  Based on this stylized fact, it is possible to assume that 
\begin{equation}
S(m_j)=2n_j m_j \approx A m_j ^{-\alpha}
\end{equation}
where $A$ is an amplification factor, and $\alpha$ is a scaling factor indicating the \textit{roughness} of the price sample. In the case where $\alpha$ is a constant, we refer to a single fractal, while if $\alpha$ is a function of persistent movement size $\alpha(m_i)$, it represents a \textit{multifractal} \cite{Mandelbrot-2001(2)}.  

The power-law assumption leads to a relationship between the expected number of persistent movements as a function of the persistent movement size.  This relationship is 
\begin{equation}
n_j = \frac{1}{2}A m_j^{-\alpha-1}
\end{equation}
Therefore, simply investigating the power-law scaling characteristics of the number of persistent movements against the persistent movement size is analogous to analyzing the contribution of each persistent movement size to the overall total variation.  

The area under the persistent movement spectra curve is representative of the persistent part of the total variation, or
\begin{equation}
\sum_{i=1}^{n_{pm}} 2 \mu_i =\sum_{j=1}^{n_{m}}  n_j m_j \approx \sum_{j=1}^{n_{m}}  A m_j^{-\alpha}
\end{equation} 
where $n_{m}$ is the number of sizes used to categorize the persistent movement sizes.

The simple interpretation of the scaling exponent is that if $\alpha$ decreases, we should expected fewer smaller size price reversals in any larger persistent price movements, and conversely if $\alpha$ increases, the price will exhibit more smaller price reversals in any given larger persistent price movements.

\section{Demonstration of the Method}

\subsection{Data Description}

\begin{figure}
	\centering 
	\includegraphics[width=10cm, height=7cm]{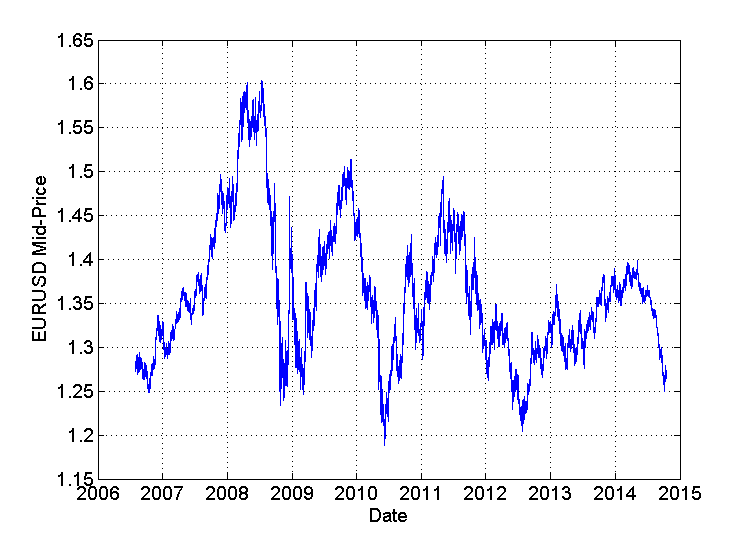}
	\caption{EURUSD Mid-Price}
	\label{eurusd}
\end{figure}

\begin{figure}
	\centering 
	\includegraphics[width=10cm, height=7cm]{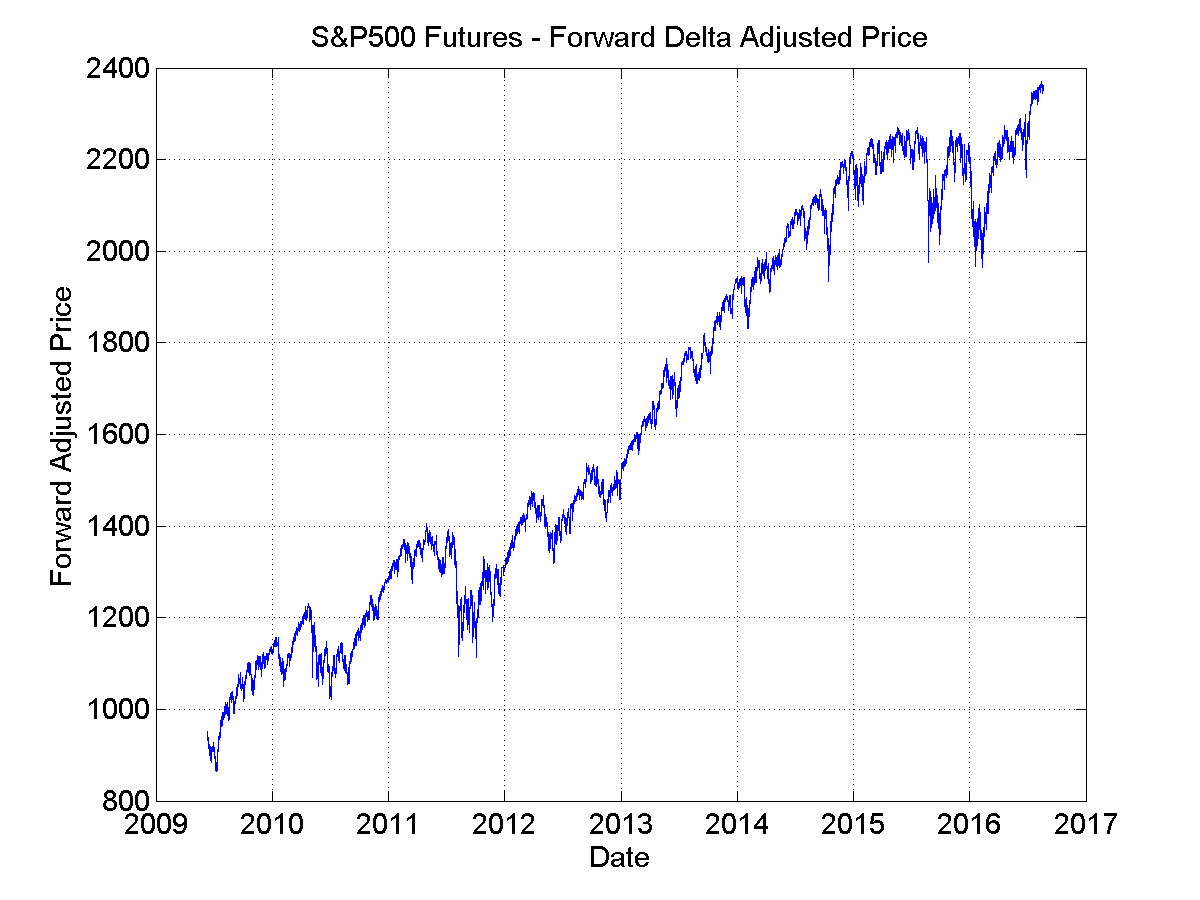}
	\caption{S\&P 500 Continuous Contract Price}
	\label{ESFUT}
\end{figure}

The analysis is  performed on price series for two of the most heavily traded financial instruments available: EURUSD spot FX and the S\&P 500 index.  For EURUSD,  The mid-price obtained from the best bid-offer data from August 2006 to October 2014 is used.  The data is plotted in Figure \ref{eurusd}. For the S\&P 500, e-Mini futures price data spanning June 2009 to August 2016 is  analyzed.  Since the data spans multiple contract expiry periods, the price sets of the selected contracts terms are merged into a differentially forward adjusted continuous contract price series.  This data maintains the price variations throughout history while removing the steps associated with rolling from one contract to the next, and hence maintains the effect of carrying charges and dividends.  The roll convention utilized for this example consists of rolling out of the current contract 6 days prior to the expiration date and into the next valid contract where only March, June, September and December contracts are considered. The resulting continuous contract price series is shown in Figure \ref{ESFUT}.  The minimum price variations, called tick sizes, are 0.0001 USD per EUR for the EURUSD spot FX data, and 0.25 USD for the S\&P 500 futures data. 

\subsection{Decomposition Results}

\begin{figure}
	\centering 
	\includegraphics[width=12cm, height=9cm]{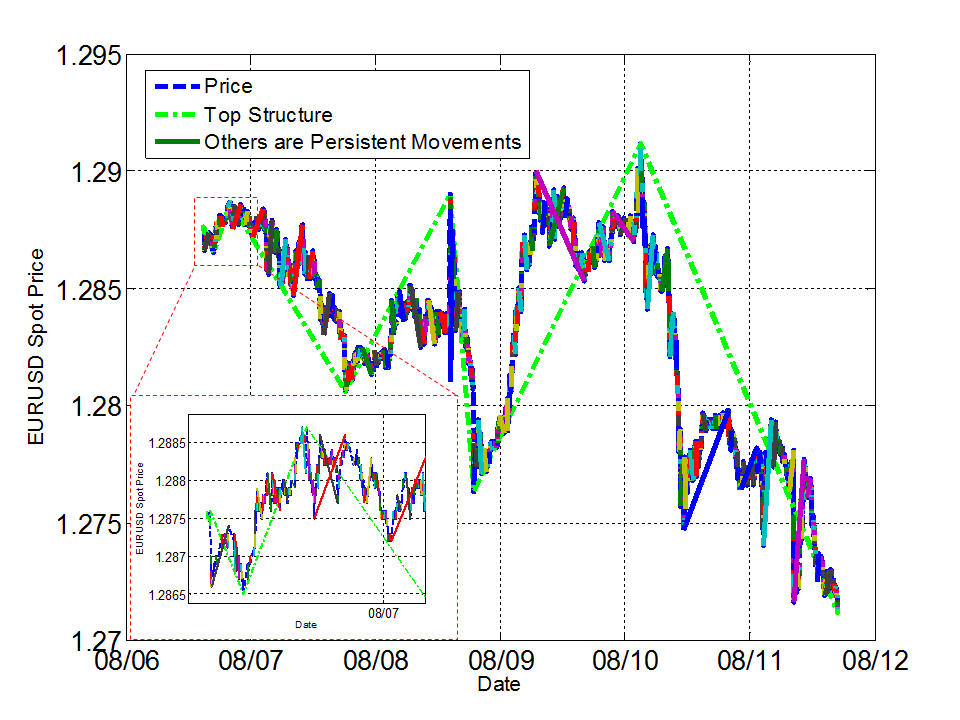}
	\caption{EURUSD Week of August 6 2006 Decomposition}
	\label{eurusd_decomp}
\end{figure}

The algorithm described in Appendix \ref{algo} is used to extract the  persistent movements for each data set.  Visualization of all persistent movements is not practical.  The decomposition covers all price reversals, from single tick-size variations to large macro movements.   It is difficult to create a good global picture in a static chart.  For demonstration, Figure \ref{eurusd_decomp} shows each component of the persistent movement decomposition of EURUSD data for the week of August 6 2006.  The non-persistent top structure is demonstrated by joining the non-persistent extrema of the price in that week.  All other parts shown in solid lines are the persistent movements displayed by joining persistent extrema with a line. This single week of data generates 7641 persistent movements with sizes ranging from 1 tick to 65 ticks.  Larger price movements observed in that week are not yet persistent and therefore, their contribution to the persistent movement spectrum is not yet reflected. The density of persistent movements in time is a natural business-time \cite{Dacorogna-2001} delineation.  

\subsection{Persistent Movement Spectra and Scaling}

\begin{figure}
	\centering 
	\includegraphics[width=15cm, height=10cm]{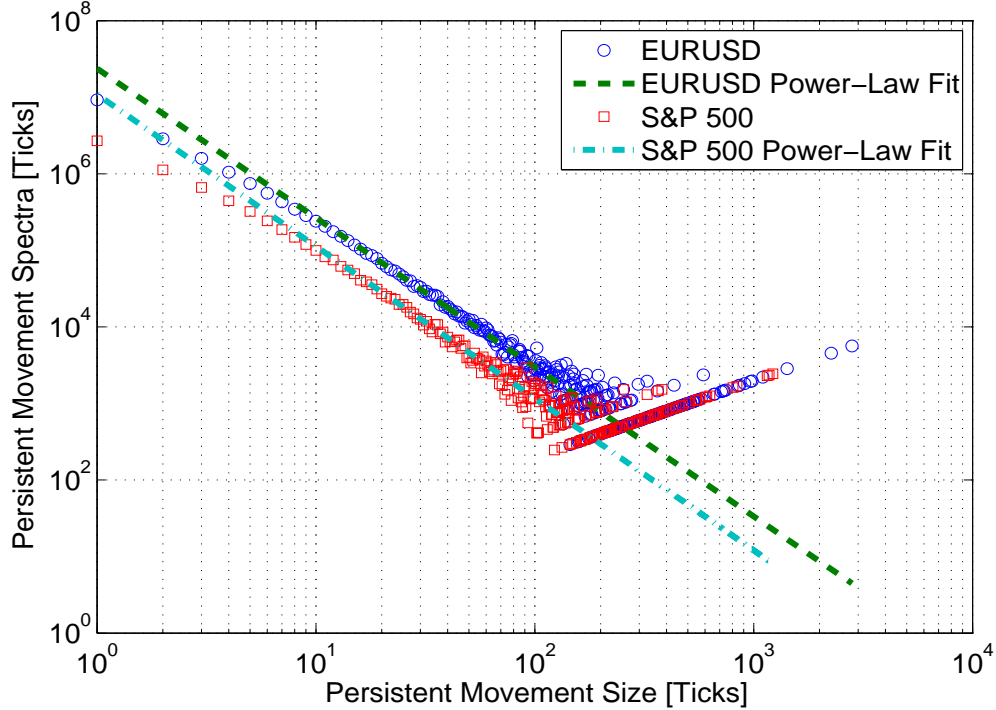}
	\caption{Persistent Movement Spectra}
	\label{pm_spectra}
\end{figure}

The persistent movement spectra results for the analysis of both data sets are shown in Figure \ref{pm_spectra}, which also includes a power law fit obtained using the maximum likelihood power-law fitting method presented by Clauset \cite{Clauset-2009}.  The fitted curve is provided up to the largest available persistent movement size in the sample.  The empirical spectra has a finite resolution dictated by the number of movement in the sample while the fitted spectra is the expected one over an infinite time horizon.   

The power-law fitting algorithm provides both a scaling exponent and a minimum price movement size for which the power law applies.  For EURUSD, the scaling exponent is estimated to $\alpha=1.97$, valid as a single fractal characteristic for persistent movement sizes higher than 11 ticks.  Smaller size variations seem to be better characterized by multifractals.  This value of scaling exponent matches the exponent obtained by Dupuis and Olsen \cite{Dupuis-2012} for the EURUSD spot FX market.  Similarly, the estimated scaling exponent of the S\&P 500 Futures price series over the complete period is $\alpha=1.985$, valid as a single fractal characteristic for persistent movement sizes higher than 13 ticks.  Again, smaller size variations are better characterized by multifractals.  The straight lines at the bottom of the actual empirical spectra are related to having only one persistent movement of that size in the data.  As a consequence, its contribution to the total variation is proportional to its size, hence forming a line.  

\begin{figure}
	\centering 
		\includegraphics[width=18cm, height=10cm]{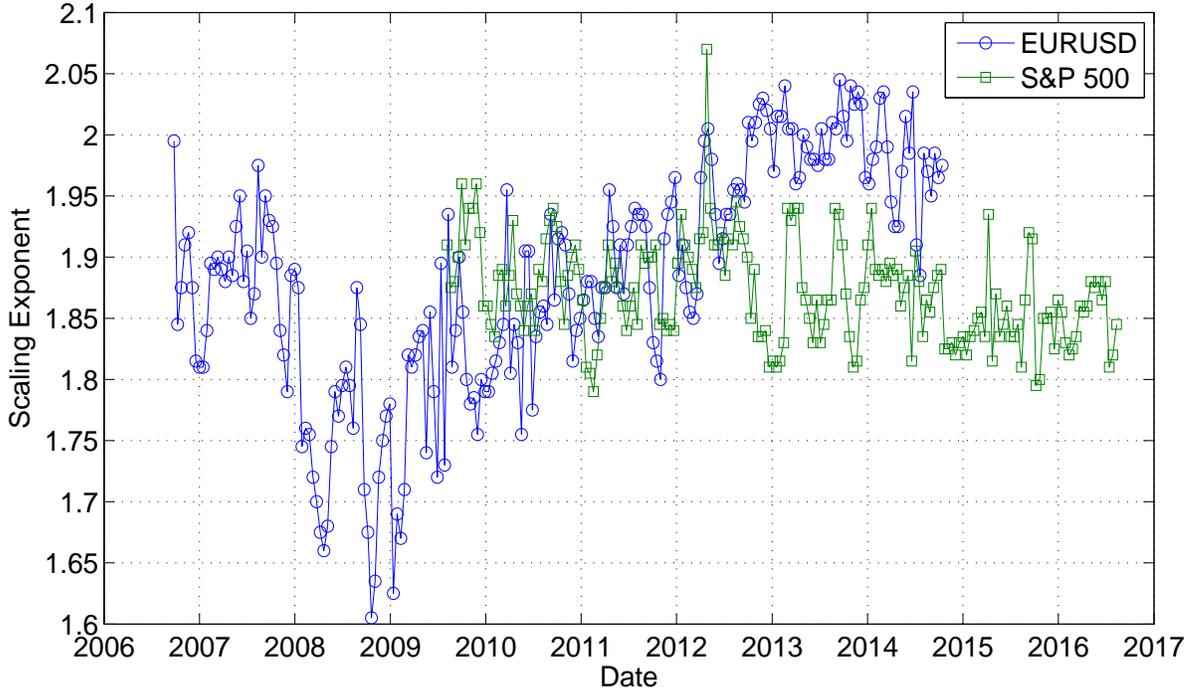}
	\caption{8-Week Recursive  Estimate of Scaling Parameter}
	\label{pm_recursive_alpha}
\end{figure}

The global estimates of the persistent movement scaling exponent provide an interesting insight into the characteristics at a high level.  But more important would be to estimate the evolution of the scaling characteristics over time.  In that case, it could be used as a local market state to inform trading decisions.  A minimum look-back period is required to ensure that enough events are observed for the estimation.  The \textit{flow} of events depends on volatility and on the minimum price variations.   Figures \ref{pm_recursive_alpha} shows recursive estimates over periods of 8 weeks and updated every second week.  These plots indicate that the scaling characteristics of sub periods are different from the scaling characteristics of the complete set they belong to.

The results show that from 2013 to 2015, one should have expected less mean reversion (a.k.a. higher scaling exponent) in the price of the S\&P 500  than in the price of EURUSD. This is obvious with hindsight by looking at the price history directly (Figures \ref{ESFUT} and \ref{eurusd}).  The scaling exponent estimation would have detected that fact in late 2012 at which time the recursive estimate of the scaling exponent for both assets diverged significantly. The scaling estimate for EURUSD also shows that in early 2008, there was a significant shift in the scaling property, a change which preceded the 2008 financial crisis, also visible in the fall of 2008.   

\section{Conclusions}
The persistent movement decomposition was used to decompose the time series of the price for two important financial instruments.  The method decomposes prices into pairs of persistent local extrema representing persistent movements with sizes summing up to its total variation.  The decomposition  maintains all price reversals, while providing a natural insight into the complex micro-structure of price series or, for that matter, any time series exhibiting fractal characteristics.   

Results of the decomposition for Spot FX EURUSD and the S\&P 500 Futures were presented. The persistent movement spectrum, representing the contribution of persistent movements of all sizes to the total variation, is shown to follow a power law.  This stylized fact about the market price evolution is indicative of the self-similarity present in complex fractional processes.  The scaling characteristics of price data support the validity of the Fractal Market Hypothesis.  According to FMH, prices evolve in response to a multitude of participants with distinct views about asset prices over various time horizons.  The investigations of the scaling characteristics of an asset price can therefore be interpreted as a way to detect the distribution of such participants, and could potentially be used to inform trading. 

Finally, a fundamental property of persistent movement decomposition not discussed in this paper is the tree structure of persistent movements whereby smaller components are contained in larger ones.  The properties of the resulting tree structure, indicative of the evolution of dependence between movements at various sizes, contains further valuable information about markets.

\bibliographystyle{ieeetr}

\bibliography{biblio}

\begin{thebibliography}{10}

\bibitem{Elliot-1938}
R.~Elliot, {\em The Wave Principle}.
\newblock BN Publishing, 2012.

\bibitem{Mandelbrot-1963}
B.~Mandelbrot, ``The variation of certain speculative prices,'' {\em The
  Journal of Business}, vol.~36, pp.~394--419, October 1963.

\bibitem{Mandelbrot-1997}
B.~Mandelbrot, {\em Fractals and Scaling in Finance}.
\newblock Springer Verlag Science, 1~ed., 1997.

\bibitem{Mandelbrot-2001(1)}
B.~Mandelbrot, ``Scaling in financial prices: {I.} tails and dependence,'' {\em
  Quantitative Finance}, vol.~1, pp.~113--123, 2001.

\bibitem{Mandelbrot-2001(2)}
B.~Mandelbrot, ``Scaling in financial prices: {II.} multifractals and the star
  equation,'' {\em Quantitative Finance}, vol.~1, pp.~124--130, 2001.

\bibitem{Mandelbrot-2001(3)}
B.~Mandelbrot, ``Scaling in financial prices: {III.} cartoon brownian motions
  in multifractal time,'' {\em Quantitative Finance}, vol.~1, pp.~427--440,
  2001.

\bibitem{Muller-1990}
U.~M\"{u}ller, M.~Dacorogna, R.~Olsen, O.~Pictet, M.~Schwarz, and C.~Morgenegg,
  ``Statistical study of foreign exchange rates, empirical evidence of a price
  change scaling law, and intraday analysis,'' {\em Journal of Banking and
  Finance}, vol.~14, pp.~1189--1208, 1990.

\bibitem{Bouchaud-2000}
J.-P. Bouchaud, ``Power-laws in economy and finance: some ideas from physics,''
  {\em arXiv:cond-mat/0008103}, August 2000.

\bibitem{DiMatteo-2003}
T.~DiMatteo, T.~Aste, and M.~Dacorogna, ``Scaling behaviors in differently
  developed markets,'' {\em Physica A}, pp.~183--188, 2003.

\bibitem{Gencay-2001}
R.~Gencay, F.~Selcuk, and B.~Whitcher, ``Scaling properties of foreign exchange
  volatility,'' {\em Physica A}, pp.~249--266, 2001.

\bibitem{Xu-2003}
Z.~Xu and R.~Gencay, ``Scaling, self-similarity and multifractality in fx
  markets,'' {\em Physica A}, pp.~578--590, 2013.

\bibitem{Peters-1997}
E.~Peters, {\em The Fractal Market Hypothesis}.
\newblock Wiley, 1~ed., 1994.

\bibitem{Kristoufek-2012}
L.~Kristoufek, ``Fractal market hypothesis and global financial crisis,'' {\em
  arXiv:1203.4979v1}, March 2012.

\bibitem{Hurst-1951}
H.~Hurst, ``Long-term storage capacity of reservoirs,'' {\em Transactions of
  the American Society of Civil Engineers}, vol.~1, pp.~770--799, 1951.

\bibitem{Lo-1991}
A.~W. Lo, ``Long-term memory in stock market prices,'' {\em Econometrica},
  vol.~59, no.~5, pp.~1279--1313, 1991.

\bibitem{Dupuis-2012}
A.~Dupuis and R.~Olsen, {\em High Frequency Finance: Using Scaling Laws to
  Build Trading Models}, ch.~20, pp.~563--582.
\newblock Wiley and Son, 2012.

\bibitem{Machado-2102}
J.~Machado and F.~Duarte, ``Analysis of financial indices by means of the
  windowed fourier transform,'' {\em Signal Image and Video Processing},
  vol.~6, pp.~487--494, September 2012.

\bibitem{Machado-2011}
J.~Machado, F.~Duarte, and G.~Duarte, ``Analysis of financial data series using
  fractional fourier transform and multidimensional scaling,'' {\em Nonlinear
  Dynamics}, vol.~65, pp.~235--245, August 2011.

\bibitem{Bayraktar-2008}
E.~Bayraktar, H.~Poor, and K.~Sircar, ``Estimating the fractal dimension of the
  sp500 index using wavelet analysis,'' {\em arXiv:math/0703834v1}, February
  2008.

\bibitem{Lemire-2009}
D.~Lemire, M.~Brooks, and Y.~Yan, ``An optimal linear time algorithm for
  quasi-monotonic segmentation,'' {\em International Journal of Computer
  Mathematics}, vol.~86, no.~7, pp.~1093--1104, 2009.

\bibitem{Brooks-2016}
M.~Brooks, ``Varilets: Additive decomposition, topological total variation, and
  filtering of scalar fields,'' {\em arXiv:1503.04867v3}, April 2016.

\bibitem{seversky-2016}
L.~M. Seversky, S.~Davis, and M.~Berger, ``On time-series topological data
  analysis: New data and opportunities,'' in {\em Proceedings of the IEEE
  Conference on Computer Vision and Pattern Recognition Workshops}, pp.~59--67,
  2016.

\bibitem{Brooks-2016-2}
M.~Brooks, ``Persistence lenses: Segmentation, simplification, vectorization,
  scale space and fractal analysis of images,'' {\em arXiv:1604.07361v3}, June
  2016.

\bibitem{Edelsbrunner-2008}
H.~Edelsbrunner and J.~Harrer, ``Peristent homology - a survey,'' {\em
  Contempory Mathematics}, vol.~453, pp.~257--282, 2008.

\bibitem{Lemire-2005}
D.~Lemire, M.~Brooks, and Y.~Yan, ``An optimal linear time algorithm for
  quasi-monotonic segmentation,'' in {\em Proceedings of the Fifth IEEE
  International Conference on Data Mining (ICDM-05)}, pp.~709--712, IEEE, 2005.

\bibitem{Dacorogna-2001}
M.~Dacorogna, R.~Gen\c{c}ay, U.~M\"{u}ller, R.~Olsen, and O.~Pictet, {\em An
  Introduction to High-Frequency trading}.
\newblock Academic Press, 1~ed., 2001.

\bibitem{Clauset-2009}
A.~Clauset, C.~Rohilla~Shalizi, and M.~Newman, ``Power-law distributions in
  empirical data,'' {\em SIAM Rev.}, vol.~51, pp.~661--703, 2009.

\end{thebibliography}

\appendices

\section{Persistent Pairs Detection Algorithm}
\label{algo}

The algorithm to extract persistent pairs from time series was taken from Lemire \cite{Lemire-2009} \cite{Lemire-2005} and is shown here.  It detects the persistent pairs \textit{on the fly} and generates a list of persistent pairs $pp$ as well as a top structure $Top$.
\\
\small
\begin{tabbing}
	AA \= AA \= AA \= AA \= AA \= AA \= \kill 
	$\bullet$ \textbf{Initialization} \\
	$firstSample=true$ \\
	$prevDirection = 0$ \\ 
	Define a dynamic array $Top$ (to contain extrema of non-yet persistent elements) \\ 
	Define an output array $pp$ (to contain extrema of persistent pairs) \\
	\\
	\\
	
	$\bullet$ \textbf{Extracting persistent pairs }\\
	For $i=1 \ldots n$ (scan each point in the time series) \\
	\> If ($firstSample=true$) \\
	\> \> \# handle initialization \\
	\> \> $firstSample=false$ \\
	\> \> $time = t_i$ \\
	\> \> $sample = x_i$ \\
	\> Else if $\lvert sample-x_i \lvert <\epsilon$ \\
	\> \> \#Update time only - same sample \\
	\> \> $time = t_i$ \\
	\> Else (New sample to be processed) \\
	\> \> \# get and process the direction of current variation \\
	\> \> $d_i = sign(x_i-sample)$ \\
	\> \> If ($d_i*prevDirection \leq 0$) \\
	\> \> \> $prevDirection  = d_i$ \\
	\> \> \> $push \{time,sample\} \rightarrow	Top $ \\
	\> \> \> If ($d_i>0$) \\
	\> \> \> \> If ($topIndex_{min}==-1$) \\
	\> \> \> \> \> $topIndex_{min} = size(T)$ \\
	\> \> \> Else if ($d_i<0$) \\
	\> \> \> \> If ($topIndex_{max}==-1$) \\
	\> \> \> \> \> $topIndex_{max} = size(T)$ \\ \\
	\> \> $time = t_i$ \\
	\> \> $sample = x_i$ \\
	\\
	\> \> \# Check for new persistent pairs \\
	\> \> $loopOn = true$\\
	\> \> while ($loopOn$) \\
	\> \> \> $lastIndex = size(Top)$ \\
	\> \> \> $prevIndex = lastIndex-1$ \\
	\> \> \> If ($lastIndex = >0$) \\
	\> \> \> \> $x_{ext} = Top(prevIndex).sample$ \\
	\> \> \> If ($d_i>0$) \\
	\> \> \> \> If ($topIndex_{max}>-1$) \\
	\> \> \> \> \> If $(topIndex_{max}=prevIndex) and (x_i\geq x_{ext})$ \\
	\> \> \> \> \> \> $topIndex_{max}=-1$ \\
	\> \> \> \> \> Endif \\
	\> \> \> \> Endif \\
	\> \> \> \> If $(topIndex_{min}=lastIndex) and (p_i < x_{ext})$ \\
	\> \> \> \> \> $loopOn=false$ \\
	\> \> \> \> Endif \\
	
	\> \> \> ElseIf ($d_i<0$) \\
	\> \> \> \> If ($topIndex_{min}>-1$) \\
	\> \> \> \> \> If $(topIndex_{min}=prevIndex) and (x_i\leq x_{ext})$ \\
	\> \> \> \> \> \> $topIndex_{min}=-1$ \\
	\> \> \> \> \> Endif \\
	\> \> \> \> Endif \\
	\> \> \> \> If $(topIndex_{max}=lastIndex) and (p_i > x_{ext})$ \\
	\> \> \> \> \> $loopOn=false$ \\
	\> \> \> \> Endif \\
	\> \> \> If ($loopOn$) \\
	\> \> \> \> \# Persistent Pair Detected \\
	\> \> \> \> $pp(ppCount) = \{Top(lastIndex), Top(prevIndex)\}$ \\
	\> \> \> \> $ppCount = ppCount + 1$ (increment persistent pair counter) \\
	\> \> \> \> pop $Top$ twice  (remove last 2 entry in array) \\
	\> \> \> Endif \\
	\> \> Endwhile \\
	\> Endif \\
	Endfor 
	
\end{tabbing}
\normalsize

\end{document}